\pdfoutput=1
\documentclass[runningheads]{llncs}

\usepackage[nocompress]{cite}
 \usepackage{amsmath} 
 \usepackage{amssymb}
 \usepackage{bm}
\usepackage{wrapfig} 
 \usepackage{booktabs}

 \usepackage{graphicx}
\usepackage{caption}
\usepackage{subcaption}
\usepackage{multirow}
\usepackage[T1]{fontenc}
 \usepackage{booktabs}
 \usepackage{threeparttable}
\usepackage{hyperref}
\usepackage{cleveref}
\hypersetup{hidelinks,
	colorlinks=true,
	allcolors=black,
	pdfstartview=Fit,
	breaklinks=true}

\usepackage{graphicx}
\begin{document}

\title{SP-GCRL: Influence Maximization on Incomplete Social Graphs}

\author{Haohua Niu\inst{1}\thanks{Haohua Niu and Yuxuan Yang share the same contribution} \and
Yuxuan Yang\inst{2}  \and
Lingfeng Zhang\inst{3} \and 
Hao Li\inst{4} \and
Jiao Liang\inst{1} \and
Zongfu Luo\inst{1}\thanks{Corresponding author} \and
Luca Rossi\inst{3}\thanks{Corresponding author}}

\authorrunning{H Niu et al.}
\institute{School of Systems Science and Engineering, Sun Yat-sen University, Guangzhou, China\\
\email{luozf@mail.sysu.edu.cn}\and
School of Finance, Jiangxi University of Finance and Economics, Jiangxi, China \and
Department of Electrical and Electronic Engineering, The Hong Kong Polytechnic University, Hong Kong, China\\
\email{luca.rossi@polyu.edu.hk}\and
School of Computer Science and Engineering, Anhui University of Science and Technology, Anhui, China
}

\maketitle              % typeset the header of the contribution
\begin{abstract}
Influence maximization (IM) in real platforms is challenged by incomplete, noisy social graphs and non-stationary diffusion dynamics. We propose SP-GCRL, a social-propagation–aware graph contrastive reinforcement learning framework that learns end-to-end seed selection under partial observability.
%Influence maximization (IM) aims to select seed nodes under a budget $k$ to maximize expected diffusion, with applications in public opinion management, public health, and marketing. In practice, platform privacy constraints and partially open APIs render observed graphs incomplete and noisy; under such conditions, classical approximations based on fixed diffusion rates and submodular monotonicity struggle to capture the nonlinear and drifting nature of real social propagation. To address this, we propose the SP-GCRL framework:
We first introduce a social-propagation-aware nonlinear diffusion function to model reinforcement/diminishing effects and probability drift under repeated exposure; we then construct dual structural views and perform contrastive learning to obtain node representations robust to missing edges and weak ties, while replacing expensive strategy metrics with a GAT-based regression surrogate to improve efficiency and scalability; finally, we use DDQN to learn an end-to-end seed selection policy on top of these representations. Experiments on multiple real-world networks show that SP-GCRL achieves significant gains over heuristic and learning-based baselines across budgets and topologies, while maintaining strong large-scale scalability. Code is available at \href{https://github.com/gcrlgraph/SP-GCRL}{here}

\keywords{Social Networks  \and Influence Maximization}
\end{abstract}
\section{Introduction}

Social networks have transformed interpersonal communication and the diffusion of information at scale, yielding rich observational traces for modeling user behavior and preferences. Within this landscape, Influence Maximization (IM) is a foundational problem with broad applications in viral marketing\cite{chen2010scalable}, recommender systems\cite{ye2012exploring}, and policy intervention\cite{klages2022optimal}: given a budget $k$, select a seed set of $k$ users that maximizes the expected cascade size under a predefined diffusion mechanism. The identification and selection of users with global influence can substantially augment the efficacy and reach of social campaigns and targeted interventions.\cite{kempe2003maximizing}.

The classical line of work initiated by\cite{kempe2003maximizing} formalizes IM under the Independent Cascade (IC) and Linear Threshold (LT) models, proving NP-hardness while establishing a $(1-1/e)$-approximation via a greedy algorithm that exploits the submodularity of the expected spread. Building on this theory, a large literature has refined scalable heuristics and sampling-based estimators for IM\cite{leskovec2007cost, tang2014influence, tang2015influence}. Yet, real social systems exhibit heterogeneous and context-dependent communication patterns that deviate from these stylized models. Empirical studies document mechanisms such as selective exposure and homophily that induce systematic biases in who is exposed and who adopts\cite{bail2018exposure}, while communication unfolds through multi-stage, feedback-coupled dynamics shaped by network structure and context\cite{aiyappa2024emergence}. These observations motivate diffusion models that are both more expressive and more robust to data imperfections.

Recent learning-based approaches address part of this need by leveraging Graph Neural Networks (GNNs) and Reinforcement Learning (RL) to directly approximate influence dynamics from data\cite{kumar2022influence, li2022piano, chen2023touplegdd, chen2024social}. However, in many practical settings the complete network and diffusion histories are not observable. Privacy constraints, API limitations, sampling bias, missing or delayed logs, and incomplete cross-platform linkages lead to partially observed graphs and cascades, with unobserved edges, unrecorded exposures, and censored adoptions. Such incompleteness fundamentally challenges both parametric diffusion modeling and representation learning, thus degrading IM performance.% if not explicitly addressed.

To address these challenges, we propose SP-GCRL, a framework that combines self-supervised graph contrastive representation learning with RL for seed selection under partial observability. SP-GCRL learns node embeddings from incomplete graphs by contrasting structure-aware augmented views that emphasize information pathways most predictive of diffusion, and then trains a Double Deep Q-Network (DDQN) to produce seed sets from the learned representations. Our main contributions are:
\\ \indent(1) We introduce a unified probability formulation that captures both social reinforcement and social weakening, yielding a non-monotonic exposure–response curve aligned with observed diffusion behavior and accommodating deviations from IC/LT assumptions.
\\ \indent(2) We design two structural augmentations tailored to incomplete data, one that concentrates on high-information propagation corridors while being resilient to missing edges (Section~\ref{subsec:GCL}(a)) and one that emphasizes nodes with high dynamical reachability under linearized propagation (Section~\ref{subsec:GCL}(b)). Using Graph Contrastive Learning (GCL) we maximize agreement between these views producing embeddings robust to partial network and cascade observations.
\\ \indent(3) Leveraging the learned node embeddings, we train a DDQN to derive an end-to-end seed selection policy that operates effectively when only fragmentary graph or diffusion information is available, yielding the SP-GCRL framework.
\\ \indent\noindent (4) Across eight real-world datasets, under controlled masking and subsampling regimes that emulate practical incompleteness, SP-GCRL consistently outperforms heuristic and contemporary learning-based baselines in achieved influence spread and generalization.

\section{Related Work}
Research on IM spans classical diffusion modeling and learning-based algorithms. The foundational IC and LT models formalize diffusion as, respectively, independent stochastic activations and thresholded aggregations of neighbor influence\cite{kempe2003maximizing, wang2012scalable, chen2010scalable}. The triggering model unifies these paradigms by allowing node-specific triggering sets\cite{tang2015influence}. Heuristics such as PageRank are frequently employed for seed pre-selection\cite{page1999pagerank}. These models confer analytical tractability and underpin scalable IM techniques, but they rely on strong independence and stationarity assumptions that can misalign with empirical propagation, motivating methods that accommodate heterogeneous, context-dependent dynamics\cite{tang2015influence}.

To address these limitations, recent work leverages RL and GNNs. Early RL formulations cast IM as sequential decision making\cite{lin2015learning, ali2018boosting}, followed by agents that optimize seed sets via rewards from simulated cascades\cite{tian2020deep, chen2021contingency, ma2022influence} and frameworks that improve deployability through pretraining and model pools\cite{li2022piano}. In parallel, GNN-based approaches encode structural signals to predict node influence or generate diversified diffusion patterns: regression-style predictors\cite{kumar2022influence}, end-to-end deep RL with GNN encoders\cite{chen2023touplegdd}, generative IM with knowledge distillation\cite{ling2023deep}, and hybrids of graph convolution with neural bandits for exploration–exploitation under limited topology knowledge\cite{feng2024influence}, alongside scalable architectures\cite{zhu2025bigdn}. Beyond IM, GNNs have proved effective across chemistry, social analysis, and multi-agent systems\cite{du2024mmgnn, wu2022graph, zhang2024g}, motivating self-supervision. GCL constructs augmented views (e.g., node dropping, edge perturbation, attribute masking) and maximizes agreement to learn label-efficient, robust embeddings\cite{you2020graph, zhu2021empirical, hassani2020contrastive}, thereby improving generalization under noisy or incomplete observations.
\begin{figure}[h]
    \centering
    \includegraphics[width=1\linewidth]{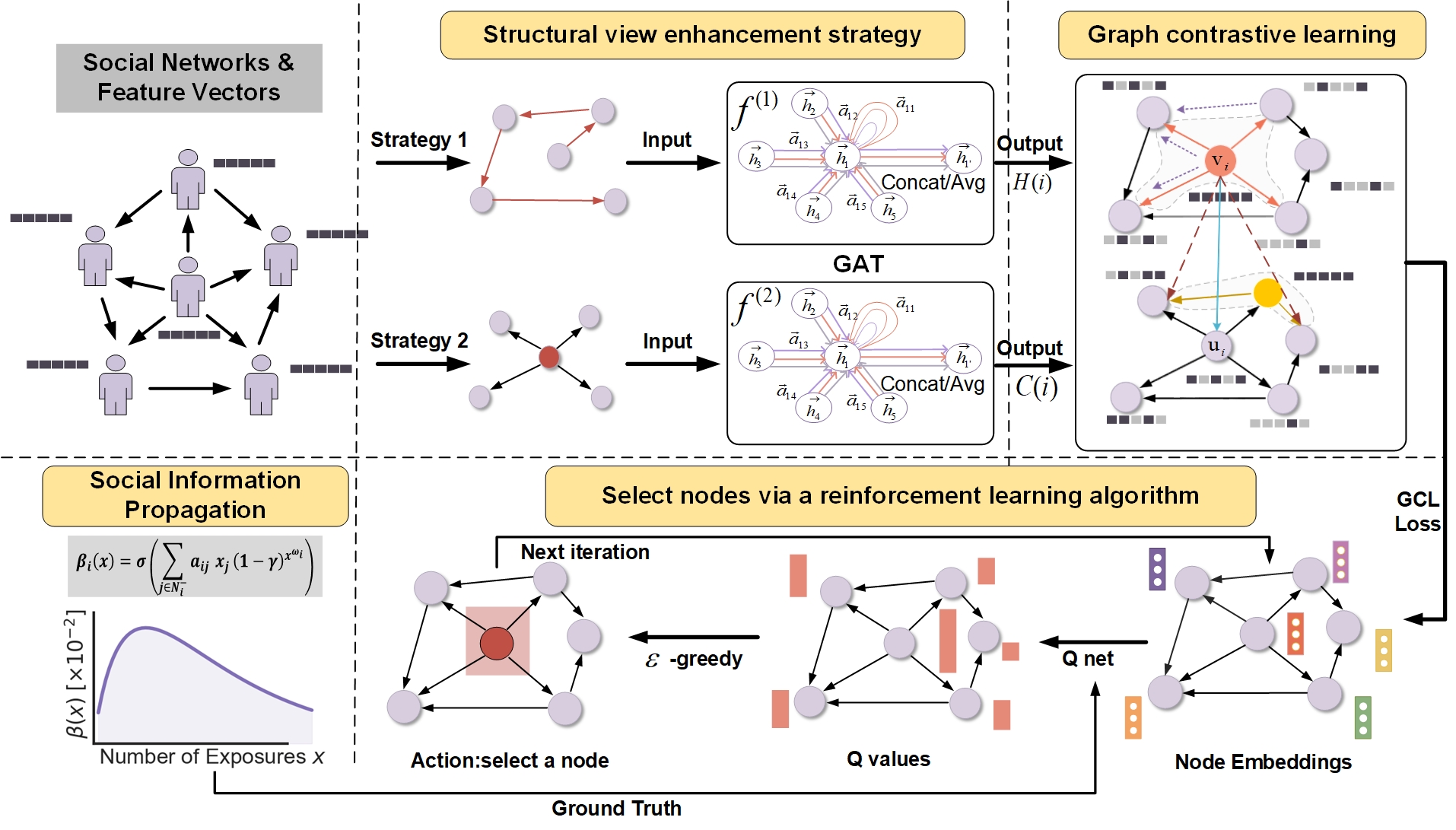}
    \caption{The SP-GCRL framework.}
    \label{fig:Influence Maximization}
\end{figure}
\section{The Proposed Framework}
\Cref{fig:Influence Maximization} presents the overall SP-GCRL pipeline. Starting from an observed (possibly incomplete) social graph, we compute an exposure-aware diffusion signal, build two structural views (backbone and controllability), learn robust node embeddings via contrastive learning with a lightweight GAT approximation, and finally select seeds using a DDQN policy to maximize expected spread. Full details are provided in the following sections.

\subsection{Social Information Propagation}

In social networks, the probability of information propagation depends on multiple complex factors. Traditional models, such as the IC model, typically assume that nodes propagate information with fixed probabilities, making it challenging to accurately capture the nonlinear effects of social relationships on user forwarding behaviors in real-world scenarios. To address this issue, inspired by the dynamic spreading equation proposed by\cite{meng2025spreading}, this study develops a nonlinear propagation model that incorporates social relationships and the ``interest threshold'' effect, which is expressed as
\begin{equation}
\beta_i(\mathbf{x})=\sum_{j\in\mathcal{N}_i^-}a_{ij}x_j\left(1-\gamma\right)^{f_i(x)} \,,
\end{equation}
where $\beta_{i}(x)$ is the probability of the $x$-th attempt of the surrounding nodes to activate the user to forward the information, $a_{i}$ represents the intrinsic spreading capability of the information for node $i$, defined as the edge weight $1/d_{in}(i)$ between nodes, where $d_{in}$ represents the in-degree of the node. $\sigma(\cdot)$ denotes the sigmoid function, $f_i(x)$ is is an exposure-adjustment factor that modulates the effective influence of repeated views by accounting for users' uncertainty about the marginal benefit of retweeting, and $\gamma$ is the average proportion of common neighbors shared between a user and its neighbors, which is calculated by the equation
\begin{equation}\label{average propagation}
\gamma=\frac1N\sum_{u=1}^N\frac1{|N_u|}\sum_{v\in N_u}\frac{|N_u\cap N_v|}{|N_u|} \,,
\end{equation}
where $N$ is the total number of nodes in the network, $N_u$ denotes the neighbor set of node $u$, and $|N_u \cap N_v|$ represents the number of common neighbors shared between nodes $u$ and $v$. The $\gamma$ in \cref{average propagation} is computed from an ego-centric viewpoint: it divides the neighbour overlap by the size of the source node’s neighbour set $|N_{u}|$ and then averages the result over all neighbours and all users.

Further, in order to determine the explicit form of the function $f_i(x)$, boundary conditions should be considered. Specifically, when a user first encounters the information ($x=1$) and decides to forward it, a proportion $\gamma$ of its neighbors are already aware of this information. Consequently, the proportion of effective new audience at this moment is $1-\gamma$. According to this boundary condition, we have $\beta_i(1)=\alpha_i(1-\gamma)$. Then, for $x > 1$, we plug $\alpha_i=\frac{\beta_i(1)}{1-\gamma}$ into \cref{average propagation} and obtain
\begin{equation}
f_i(x)=\frac{\ln\left[\frac{\beta_i(x)}{\beta_i(1)}\right]-\ln x}{\ln(1-\gamma)}+1 \,.
\end{equation}

\begin{figure}[t!]
\centering
\includegraphics[scale=0.3]{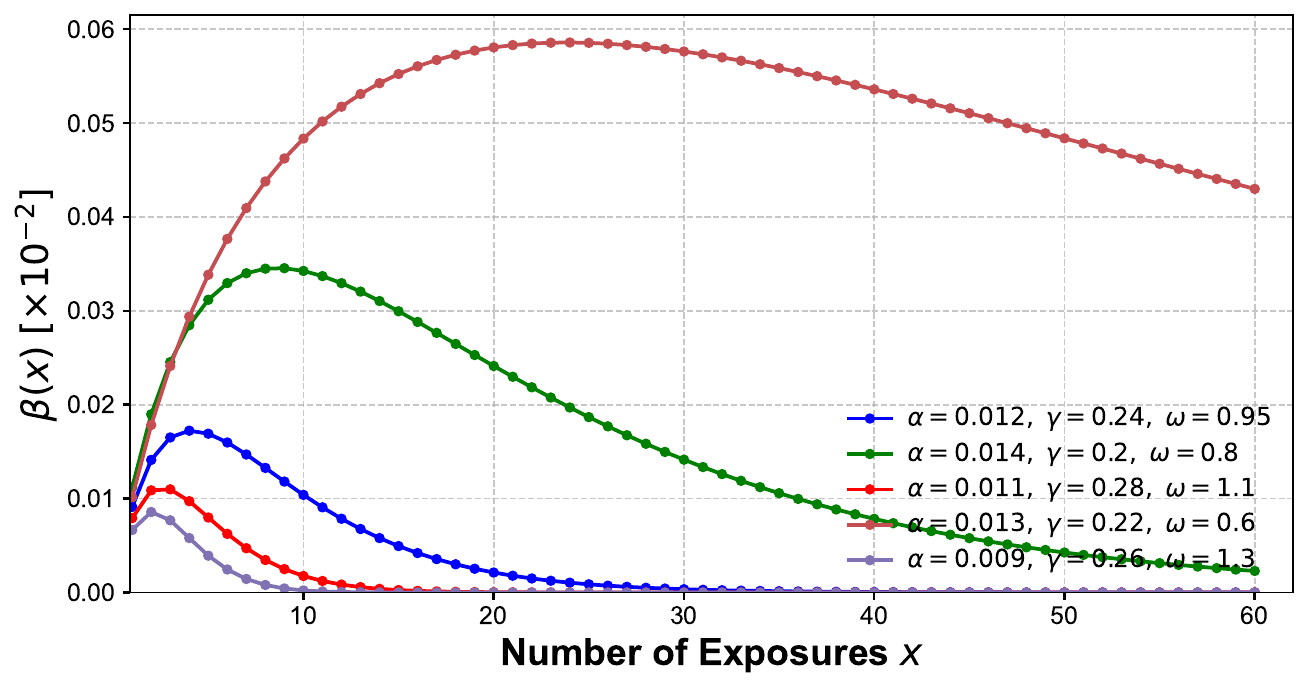}
\caption{For different choices of function parameters, the propagation probability varies as the number of exposures x increases.}
\label{fig:beta_spread_figure}
\end{figure}

Analyzing data from large-scale real-world social networks shows that the most suitable form of function to approximate $f_i(x)$ is the power-law function\cite{meng2025spreading}, expressed specifically as $f_i(x)=x^{\omega_i}$ where $\omega_i$ is a parameter reflecting users' uncertainty or deviation in evaluating forwarding benefits of the information. Therefore, the final propagation probability formula becomes
\begin{equation}
\beta_i(\mathbf{x})=\sigma\left(\sum_{j\in\mathcal{N}_i^-}a_{ij}x_j\left(1-\gamma\right)^{x^{\omega_i}}\right) \,,
\end{equation}
where the sigmoid layer $\sigma(\cdot)$ bounds the activation probability strictly within $[0,1]$ and produces a natural saturation effect, which better mirrors the empirical observation that users’ forwarding willingness plateaus after a certain number of exposures. In order to show the form of the propagation equation more intuitively, \Cref{fig:beta_spread_figure} shows how the propagation probability beta varies as the number of exposures $x$ increases, for different choices of the function parameters.

\subsection{Graph Contrastive Representation Learning} \label{subsec:GCL}
Large-scale recommendation scenarios present two major challenges, namely: 1) Label scarcity and incomplete graph structures — due to platform and policy restrictions, the visible social graph available to researchers is often only a sampled subset of the full network. In addition, privacy and access constraints frequently result in partially observed or structurally incomplete graphs, making it difficult to obtain reliable supervision signals; 2) Network noise and dynamic diffusion — weak social ties and machine accounts exist in social edges, while the propagation probability continuously drifts with both time and topic.

Traditional unsupervised methods based on reconstructed adjacency matrices or explicit probability estimation tend to rely excessively on local structure, struggle to generalize to multi-hop diffusion, or lack robustness with respect to noisy edges. To tackle these challenges, we design a GCL approach to learn node representations by maximizing the agreement of node embeddings across two subgraphs without labels. The following subsections therefore focus on the two core components of this pipeline:
1) how two types of subgraphs are constructed through topology- and feature-level perturbations, and
2) how the contrastive loss is formulated given these subgraphs.% and integrated into our RL framework.

\subsubsection{(a) Path-Entropy-Steiner Backbone Augmentation.} The propagation graph is estimated through $n$-round Monte Carlo sampling and denoted as $\mathcal{T}_{r}=(V_r,E_r)$. The path set of all nodes is defined as $\mathcal{P}_{u\to v}=\left\{p=(u,\ldots,v)\subseteq\mathcal{T}_r\right\}$, and the path probability can be obtained as $\mathrm{Pr}(p)=\prod_{(i,j)\in p}\alpha_{ij}$. From an information theoretic perspective, we define the path inverse entropy for each propagation chain $p$ as

\begin{equation}
\mathrm{H}(u,v)\mathrm{~=~}\frac1{-\sum_{p\in\mathcal{P}_{u\to v}}\Pr(p)\log\Pr(p)+\epsilon} \,.
\end{equation}

Recall that a Steiner tree in a weighted graph $G=(V,E)$ with a terminal set $T_{term}\subseteq V$ is a tree subgraph that spans all terminals and minimises the total edge cost, while being allowed to introduce additional non-terminal vertices called Steiner nodes~\cite{hwang1992steiner}. % The nodes with the top $k$ inverse entropy scores $s_v=\frac1{|V|}\sum_{u\in V}\mathrm{H}(u,v)$ are selected to form the Steiner tree node set $T$. 
The Steiner tree node set $T_{term}$ is constructed by selecting nodes with the top $k$ inverse entropy scores as
$s_v = |V|^{-1} \sum_u \mathrm{H}(u,v)$. We build a minimum Steiner tree and choose the set of edges with  minimum cost (most stable propagation) to make the set $T_{term}$ of critical nodes connected, i.e.,
\begin{equation}\label{tree}
\mathcal{B}=\arg\min_{L\subseteq E}\left\{ \sum_{i,j\in T_{term}}\mathrm{H}(i,j)\mid T_{term}\subseteq\bigcup_{(i,j)\in L}\{i,j\} \right\} \,.
\end{equation}

As the objective in~\cref{tree} is NP-hard, in practice we employ the KMB heuristic to obtain a 2-approximate Steiner backbone. Specifically, we first build a metric closure of the Steiner tree node set $T_{term}$ with edge weight $\tilde{c}_{uv}=1/\mathrm{H}(u,v)$, compute all-pairs shortest paths, and extract a minimum spanning tree in the closure. The tree is then unfolded back to the original graph, and non-terminal leaves are pruned so that the resulting backbone $\mathcal{B}$ remains sparse. The overall complexity of this procedure is $\mathcal{O}\big(|T_{term}|(m+n\log n)\big)$ in time and at most $2|T_{term}|-2$ real edges in space, where $n$ is the number of nodes and $m$ is the number of edges.

We refer to the resulting Steiner backbone as a stable graph for information diffusion. To simulate perturbations in the network structure while preserving its core connectivity, we apply noise exclusively to the non-backbone component. Specifically, we randomly eliminate a subset of edges from the non-backbone graph by sampling a random masking matrix $\tilde{{\boldsymbol{R}}}\in{0,1}^{k\times k}$, where each entry is independently drawn from a Bernoulli distribution as $\widetilde{\boldsymbol{R}}_{ij}\sim\operatorname{Bernoulli}(1-p_r)$ if $\boldsymbol{A}_{ij}=1$ for the non-backbone graph and ${\widetilde{\boldsymbol{R}}}_{ij}=0$ otherwise. Here $p_r$ is the probability of each edge being removed. The resulting adjacency matrix can be computed as
\begin{equation}
\widetilde{\boldsymbol{A}}=\boldsymbol{A}\circ\widetilde{\boldsymbol{R}} \,,
\end{equation}
where $(\boldsymbol{x}\circ\boldsymbol{y})_i=x_iy_i$ is the Hadamard product.

Apart from removing edges, we randomly mask a fraction of dimensions with zeros in node features. Formally, we first sample a random vector $\widetilde{\boldsymbol{m}}\in\{0,1\}^F$ where each dimension of it independently is drawn from a Bernoulli distribution with probability $1-p_m,\text{i.e.},\widetilde{m}_i\sim\operatorname{Bernoulli}(1-p_m),\forall i$. Then, the generated node features $\widetilde{\bm{X}}$ are computed as
\begin{equation}
\widetilde{\boldsymbol{X}}=[\boldsymbol{x}_1\circ\widetilde{\boldsymbol{m}};\boldsymbol{x}_2\circ\widetilde{\boldsymbol{m}};\cdots;\boldsymbol{x}_N\circ\widetilde{\boldsymbol{m}}]^\top \,,
\end{equation}
where $[\cdot;\cdot]$ denotes the concatenation operator. Finally, following the sequence of operations outline above, we obtain the subgraph $\tilde{G}_{1}$.

\subsubsection{(b) Gramian Control Matrix Augmentation.} In the analysis of information diffusion networks, it is essential not only to identify aggregation nodes where information converges but also to accurately locate influential ``propagation source'' nodes. These source nodes significantly enhance diffusion efficiency, enabling messages initiated from them to rapidly reach extensive areas of the network. To effectively characterize this source-driven diffusion behavior, we adopt the Gramian matrix concept from graph controllability theory. Based on this approach, we propose a subgraph generation strategy that identifies nodes exhibiting strong diffusion capabilities within the network.

We start by defining the propagation transition matrix $P$ with elements $P_{uv}=\beta_{uv}$ if $(u,v)\in E$, 0 otherwise.
We then transform $P_{uv}$ into a stochastic transition matrix byperforming the row normalization
\begin{equation}
P_{uv} = D^{-1}A,\quad D_{uu} = \sum_v A_{uv}, \quad A_{uv} = w_{uv} \,.
\end{equation}
Each element $P_{uv}$ in $P\in\mathbb{R}^{|V|\times|V|}$ represents the probability of propagating from node $u$ to $v$ in one step. Given $P$, the propagation controllability Gramian matrix of graph structure is defined as
\begin{equation}
W_c=\sum_{t=0}^JP^t(P^t)^\top \,,  
\end{equation}  
where $P^{t}$ is the $t$-th power of the propagation matrix $P$, representing the path probability of the $t$ steps propagation of between nodes and $J$ is the truncation parameter of the number of propagation steps, indicating that we only consider the information diffusion ability up to $J$ steps.

In order to accurately identify nodes with strong diffusion ability in the network, we define the propagation controllability score $C(i)$ of the nodes as the sum of the elements of row $i$ in $W_{c}$
\begin{equation}
C(i)=\sum_{j\in V}[W_c]_{ij} \,,
\end{equation}
where $[W_c]_{ij}$  indicates the total propagation energy (or influence ability) that node $i$ can exert on node $j$ over 0 to $T$ propagation steps.

Accordingly, the top $k$ nodes in terms of propagation controllability are selected to form the information diffusion source node set $\{v_{1},v_{2},...,v_{k}\}$. After determining the set of diffusion source nodes, in order to construct a structural view that is complementary to the one obtained with the Path-Entropy-Steiner Backbone Augmentation strategy (information aggregation structure), we further construct an enhanced view that can highlight the information diffusion capability of the network by centering on these source nodes. 

Finally, following the same idea as the Path-Entropy-Steiner Backbone Augmentation strategy, we perform feature masking on non-critical nodes and random edge dropping on the global graph to finally obtain the subgraph $\tilde{G}_{2}$.

\subsubsection{(c) Scaling to Large Graphs.}
On a large-scale propagation graph, computing the metrics required by the above two strategies will incur a large computational overhead. To alleviate this bottleneck, we recast the computation as a supervised regression task and employ a Graph Attention Network (GAT)~\cite{velickovic2017graph} to predict $\mathrm{H}(u,v)$ and $C(i)$, as explained below.

We start by randomly initializing the node vectors $\bm{S}_v$ and $\bm{T}_v$ to describe the ability of node $v$ to diffuse information outward when it is an ``initiator'' and its susceptibility to be activated by its neighbors when it is a ``receiver'', respectively. 

For each layer $k$ and direction $d\!\in\!\{\mathrm{out},\mathrm{in}\}$, we define the directional neighbor set
$N^d(v)$ and a shared scoring function
\begin{equation}
\psi_{u\!\to\! v}^{(k,d)}
 \;=\; \bm{a}_d^{(k)\top}\!\Big(\bm{W}_{s,d}^{(k)}\bm{S}_u^{(k)} \,\Vert\, \bm{W}_{t,d}^{(k)}\bm{T}_v^{(k)}\Big),
\end{equation}
where $\Vert$ denotes concatenation and $\bm{a}_d^{(k)}$ is trainable.
The normalized attention is the softmax over the corresponding directional neighborhood, i.e.,
\begin{equation}
\omega_{u\!\to\! v}^{(k,d)}
 \;=\; \frac{\exp\!\big(\mathrm{LeakyReLU}(\psi_{u\!\to\! v}^{(k,d)})\big)}
      {\sum\limits_{w\in N^d(v)} \exp\!\big(\mathrm{LeakyReLU}(\psi_{w\!\to\! v}^{(k,d)})\big)}.
\end{equation}
We then form a direction-aware aggregation
\begin{equation}
\bm{m}_v^{(k,d)}
 \;=\; \sum_{u\in N^d(v)} \Big(\eta_d^{(k)}\,\beta_{uv}^{(d)} \;+\; \rho_d^{(k)}\,\omega_{u\!\to\! v}^{(k,d)}\Big)\,\bm{H}_u^{(k)},
\end{equation}
where $\beta_{uv}^{(d)}$ is the diffusion prior on edge $(u,v)$ in direction $d$ (e.g., $p_{uv}$ for \textit{in}), 
$\bm{H}_u^{(k)}$ denotes the neighbor-side representation used for message passing
($\bm{H}\!=\!\bm{S}$ for $d=\mathrm{out}$, $\bm{H}\!=\!\bm{T}$ for $d=\mathrm{in}$),
and $\eta_d^{(k)},\rho_d^{(k)}$ are learnable scalars.

With this setting to hand, the node representations are updated by combining self-state and directional messages, i.e.,
\begin{equation}
\bm{S}_v^{(k+1)}
 = \sigma\!\Big(\lambda_s^{(k)} \bm{S}_v^{(k)} + \lambda_q^{(k)} \bm{m}_v^{(k,\mathrm{out})}\Big),
\bm{T}_v^{(k+1)}
 = \sigma\!\Big(\mu_t^{(k)} \bm{T}_v^{(k)} + \mu_x^{(k)} \bm{m}_v^{(k,\mathrm{in})}\Big),
\end{equation}
where $\sigma(\cdot)$ is a nonlinearity (sigmoid by default) and all $\lambda_\ast^{(k)},\mu_\ast^{(k)}$ are scalars. Finally, after $K$ layers, two MLP heads make edge-level predictions using the paired representations,
\begin{equation}
\widehat{H}(u,v)\;=\;\operatorname{MLP}_{\theta_1}\!\big([\bm{S}_u^{(K)},\bm{T}_v^{(K)}]\big),\qquad
\widehat{C}(i)\;=\;\operatorname{MLP}_{\theta_2}\!\big([\bm{S}_u^{(K)},\bm{T}_v^{(K)}]\big).
\end{equation}

\subsubsection{(d) Graph Contrastive Node Representations.} Let $\widetilde {G}_1$ and $\widetilde {G}_2$ be the two subgraphs generated using the strategies outlined above. We denote their node embeddings as $U = f(\tilde{X}_1,\tilde{A}_1)$ and $V = f(\tilde{X}_2,\tilde{A}_2)$, where $\tilde{X}_*$ and $\tilde{A}_*$ are the feature matrices and adjacency matrices of the subgraphs. 

We employ a contrastive objective (i.e., a discriminator) that distinguishes the embeddings of the same node in these two different views from other node embeddings. For any node \( v_i \), its embedding generated in one view, \( \bm{u}_i \), is treated as the anchor, its embedding generated in the other view, \( \bm{v}_i \), forms the positive sample, and embeddings of nodes other than \( v_i \) in the two views are naturally regarded as negative samples.
Formally, we define the critic \( \theta(\bm{u}, \bm{v}) = s(g(\bm{u}), g(\bm{v})) \), where \( s \) is the cosine similarity and \( g \) is a nonlinear projection to enhance the expressive power of the critic. The projection \( g \) is implemented with a two-layer MLP. Let $Q=\sum_{k=1}^N\mathbb{1}_{[k\neq i]}e^{\theta(\boldsymbol{u}_i,\boldsymbol{v}_k)/\tau}$,$Z=\sum_{k=1}^N\mathbb{1}_{[k\neq i]}e^{\theta(\boldsymbol{u}_i,\boldsymbol{u}_k)/\tau}$, We define the pairwise objective for each positive pair \( (\bm{u}_i, \bm{v}_i) \) as
\begin{equation}
\ell(\boldsymbol{u}_i,\boldsymbol{v}_i)=\log\frac{e^{\theta(\boldsymbol{u}_i,\boldsymbol{v}_i)/\tau}}{{e^{\theta(\boldsymbol{u}_i,\boldsymbol{v}_i)/\tau}}+Q+Z} \,,
\end{equation}
where $\bm{1}_{[k\neq i]}\in\{0,1\}$ is an indicator function that equals to 1 if $k\neq i$, and $\tau$ is a temperature parameter. The overall objective to be maximized is then defined as the average over all positive pairs, formally given by
\begin{equation}
\mathcal{J}=\frac1{2N}\sum_{i=1}^N\left[\ell(\boldsymbol{u}_i,\boldsymbol{v}_i)+\ell(\boldsymbol{v}_i,\boldsymbol{u}_i)\right] \,.
\end{equation}

\subsection{Reinforcement Learning}

Based on the obtained node embeddings, the score function
to measure the marginal gain of a node $u\in\bar{\mathcal{S}}_t=V\setminus \mathcal{S}_t$ with respect to the current seed set $\mathcal{S}_{t}$ is defined as
\begin{equation}
\hat{Q}(u,\mathcal{S}_t;\Theta)=\theta_1^\top\mathrm{ReLU}(\theta_2z_u^{(K)}) \,,
\end{equation}
where $\Theta = (\theta_1,\theta_2)$ are model parameters and $z_u$ is the node embedding generated by GCL. We train all the parameters end-to-end using RL. %If we denote by $\mathbb{Q}^{*}$ the optimal $\text{Q}$-function for this RL problem, then our embedding parameterized function $\hat{Q}(u,\hat{\mathcal{S}}_t;\Theta)$ will be a function approximator for it, which will be learned by DDQN.

Specifically, we use a DDQN to perform end-to-end learning of the parameters in $\hat{Q}(u_{t},\mathcal{S}_{t};\Theta)$, as this allows us to avoid the over-optimistic issue of a simple DQN by adopting two networks, a behavior network and a target network, parameterized with $\Theta$ and $\Theta^{\prime}$, respectively. The target network provides $\text{Q}$-values estimation of future states during training of the behavior network, and only updates parameters $\Theta^{\prime}$ from the behavior network $\Theta$ every episodes. We use the term episode to represent a complete sequence of node additions starting from an empty set until termination, and a single action (node addition) within an episode is referred to as a step. To collect a more accurate estimate of future rewards, n-step Q-learning is utilized to update the parameters, which is to wait for n steps before updating parameters. Additionally, we apply the fit Q-iteration with experience replay for faster learning convergence. Formally, the update is performed by minimizing the square loss
\begin{equation}
(y-\hat{Q}(u_t,\mathcal{S}_t;\Theta))^2 \,,
\end{equation}
where $y=\sum_{i=0}^{n-1}\rho^ir(\mathcal{S}_{t+i},u_{t+i})+\rho^n\max_v\:\hat{Q}(v,\mathcal{S}_{t+n};\Theta^{\prime})$, and $\rho\in[0,1]$ is the discount rate, determining the importance of future rewards.

\section{Experiments}

\subsection{Experiment Setup}
The proposed SP-GCRL is compared with other approaches over eight real-world datasets, including Petster-hamster\cite{kunegis2013konect}, Tv-show\cite{nr}, Politician\cite{nr}, Advogato\cite{nr}, Public\cite{nr}, Epinions\cite{kunegis2013konect}, Twitter\cite{feng2024influence} and Weibo\cite{feng2024influence}. We also adopted a mini-dataset radoslaw-email\cite{nr} with 167 nodes to provide a qualitative evaluation of a case study. See Table \ref{tab:dataset-statistics} for a summary of dataset statistics . We evaluate our method against a range of representative baselines for influence maximization, encompassing random selection, traditional heuristics, and learning-based approaches. The traditional baseline used is PageRank\cite{page1999pagerank}, while learning-driven models comprise gIM\cite{shahrouz2021gim}, S2V-DQN\cite{khalil2017learning}, ToupleGDD\cite{chen2023touplegdd}, DeepIM\cite{chen2024social}, and BIGDN\cite{zhu2025bigdn}.
\begin{table}[h!]
  \centering
  \begin{threeparttable}
  \caption{Statistics of the datasets used in this study.}
  \label{tab:dataset-statistics}
  \fontsize{7}{8}\selectfont
  \setlength{\tabcolsep}{6pt}          % 列间距；可调 4–6
  \renewcommand{\arraystretch}{0.9}   % 行距
  \begin{tabular}{@{} l l r r @{}}
    \toprule
    \textbf{Category} & \textbf{Dataset} & \textbf{Nodes} & \textbf{Edges} \\
    \midrule
    Social network & Petster-hamster                & 921    & 4,032   \\
    Social media pages & Tv-show                        & 3,892   & 17,262  \\
    Social media pages & Politician                     & 5,908   & 41,729  \\
    Trust network & Advogato                       & 6,551   & 51,317  \\
    Social network & Public                         & 11,565  & 67,114  \\
    Review network & Epinions                       & 26,588  & 100,120 \\
    \addlinespace[2pt]
    Social media & Twitter                        & 11.6M  & 341.8M \\
    Social media & Weibo                          & 1.8M   & 23.8M  \\
    Organizational email network & radoslaw-email\textsuperscript{\dag} & 167 & 4,250   \\
    \bottomrule
  \end{tabular}
  \begin{tablenotes}[flushleft]
    \item \footnotesize \textsuperscript{\dag}For data visualization experiments only
  \end{tablenotes}
  \end{threeparttable}
\end{table}

Edge weights on validation datasets and testing datasets have the same setting. To simulate incompletely observed networks, we process all datasets as follows: we randomly drop 50\% of the edges and mask 50\% of the node features to capture missing links and partially observed/noisy attributes. All models are evaluated under this setting. For each testing dataset, we vary the budget $b$ such that $b\in\{10,20,30,40,50\}$. All experiments are conducted on a machine with an Intel(R) Xeon(R) Platinum 8350C CPU (2.60 GHz, 48 cores), 384 GB of DDR4 RAM, an Nvidia RTX 4090 GPU with 24-GB memory, and running Ubuntu 20.04.

\subsection{Propagation Equation Fitting}
\begin{figure*}[h]
  \centering
    \subfloat[Petster-hamster]{\includegraphics[width=0.26\textwidth]{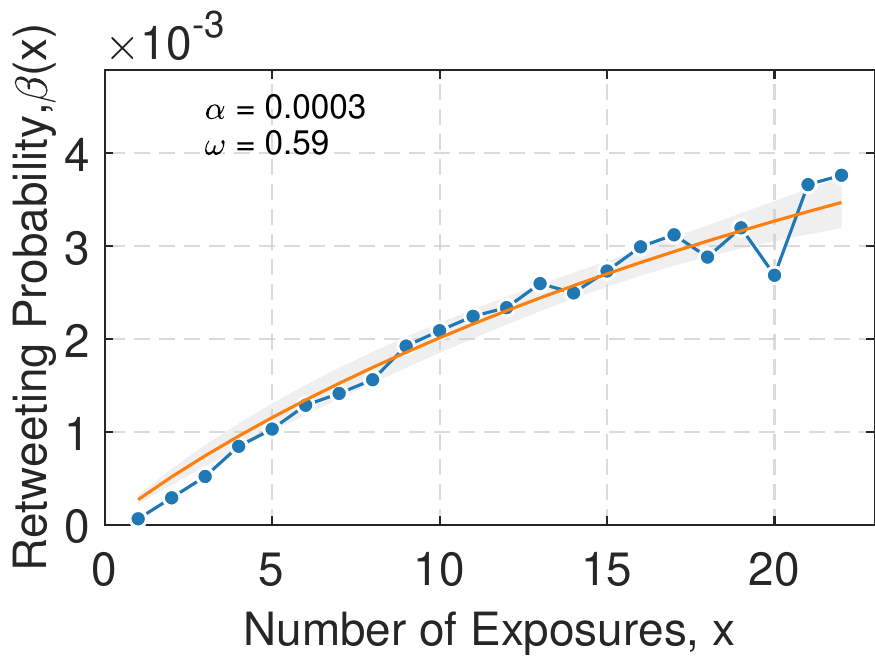}}\hspace{5pt}
    \subfloat[Tv-show]{\includegraphics[width=0.26\textwidth]{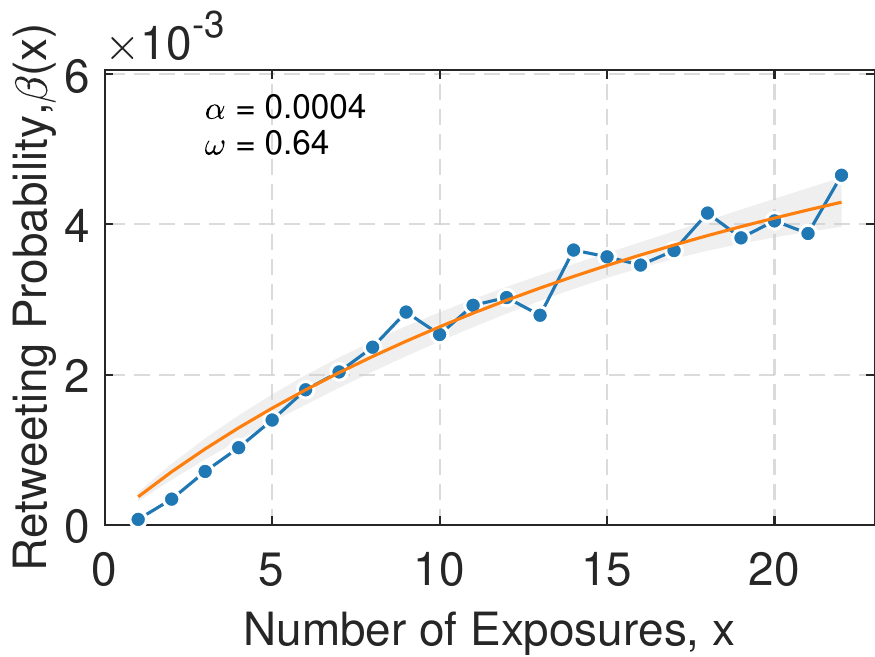}}\hspace{5pt}
    \subfloat[Politician]{\includegraphics[width=0.26\textwidth]{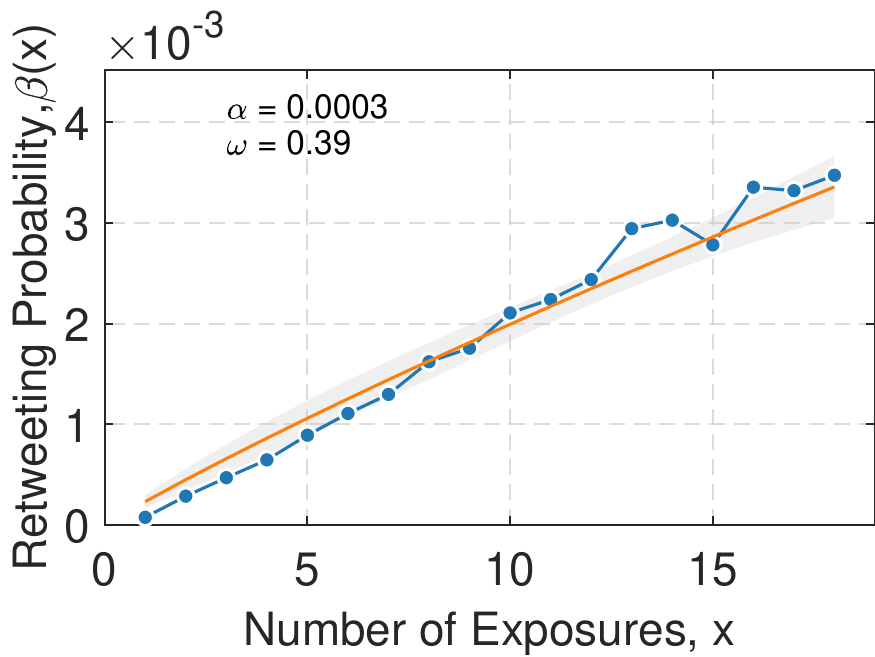}}\\[5pt]
    
    \subfloat[Advogato]{\includegraphics[width=0.26\textwidth]{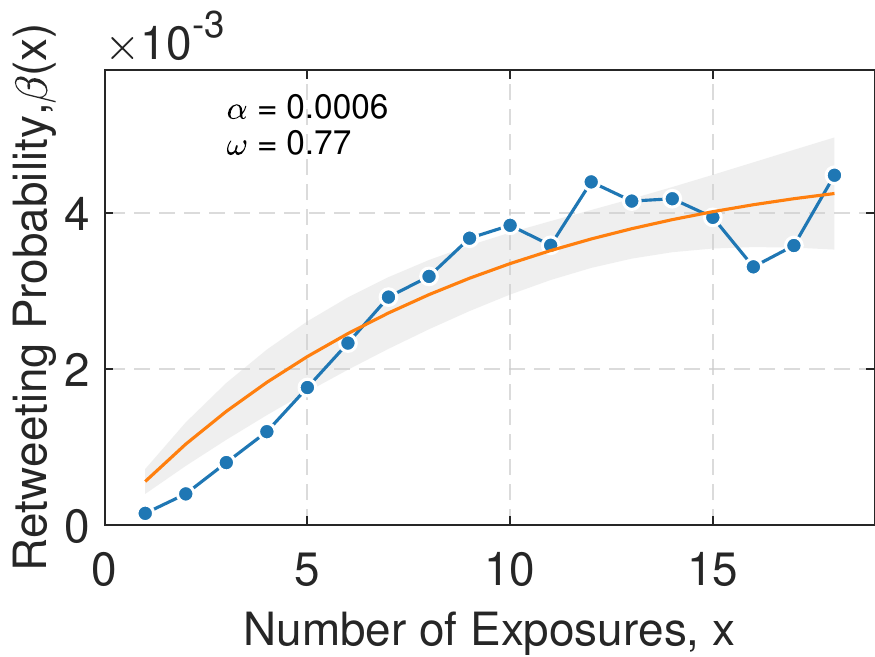}}\hspace{5pt}
    \subfloat[Public]{\includegraphics[width=0.26\textwidth]{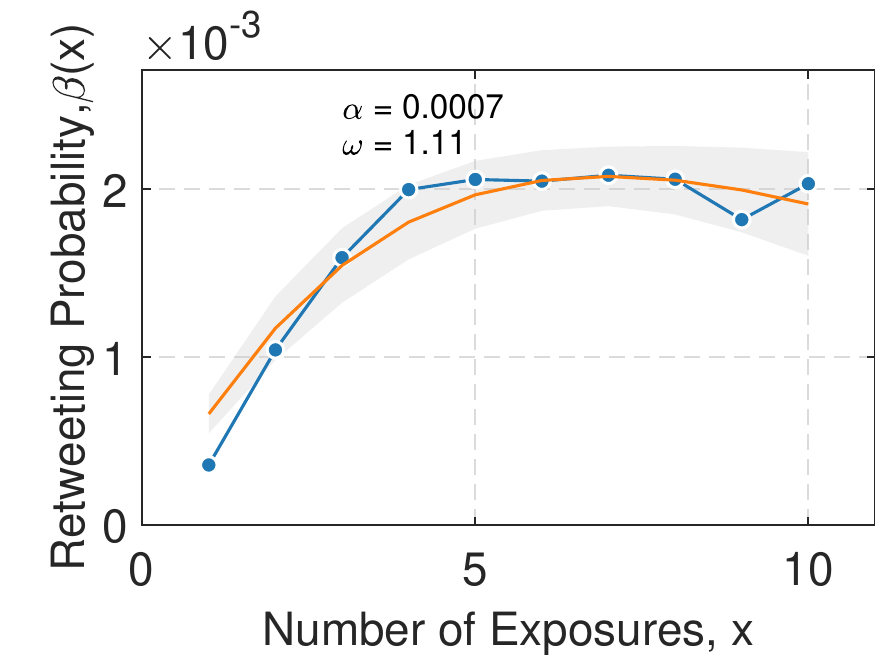}}\hspace{5pt}
    \subfloat[Epinions]{\includegraphics[width=0.26\textwidth]{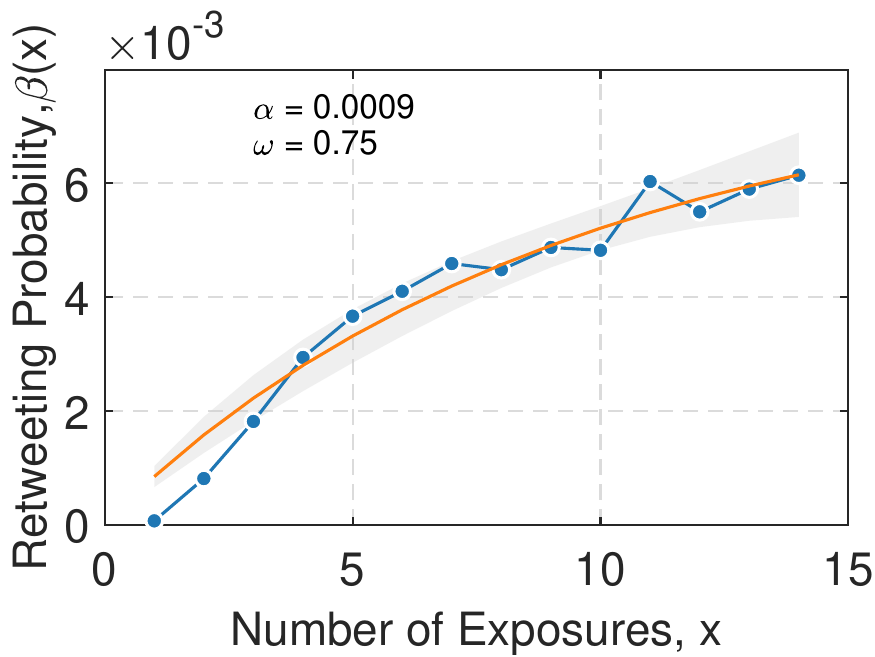}}

    \caption{Fitting of the exposure--response curve on six datasets.}
    \label{fig:fitting}
\end{figure*}

We fit the proposed social propagation equation on six datasets and find that it captures both global trends and local fluctuations of retweet probability with respect to exposure. As shown in \Cref{fig:fitting} (a)–(f), the model closely matches empirical traces within the 95\% confidence intervals, indicating robustness to sampling noise. Parameter analysis shows that $\alpha$ increases from 0.0003 to 0.0009, reflecting substantial variation in baseline diffusion strength, while $\omega$ ranges from 0.59 to 1.11, revealing heterogeneous exposure sensitivity. When $\omega \leq 1$, marginal gains from additional exposures diminish, whereas for $\omega > 1$ (\Cref{fig:fitting} (e)), saturation occurs near $x \approx 6$, suggesting information overload. Overall, the equation provides accurate fitting across diverse networks while offering interpretable parameters for downstream analysis.

\subsection{Influence Diffusion}
Table \ref{tab:performance-framework-all3} reports the expected influence spread achieved by all competing methods under budgets ranging from 10 to 50 on six real-world networks. Across all datasets and budgets, SP-GCRL achieves the highest influence spread and maintains a consistent lead over all baselines. On the small and medium graphs in Table I, the advantage is steady on Petster-hamster and Tv-show and expands with the seed budget; the gap becomes pronounced on Politician, indicating that contrastive representation learning benefits decision quality under sparse and heterogeneous topologies. Table II shows that the lead remains on Advogato, Public, and Epinions, where spreads are higher overall and fluctuate across budgets, yet SP-GCRL dominates most operating points, reflecting stable gains under differing graph densities and community structures. Table III further demonstrates scalability: on the large-scale Twitter and Weibo networks, SP-GCRL consistently ranks first for every budget, and the separation from the next best methods is maintained or amplified, suggesting that the learned local-to-global invariances improve marginal-gain estimation at scale. Taken together, the results support that coupling GCL with DDQN yields robust, budget-amplified improvements across networks of varying sizes and connectivity patterns.
\begin{table*}[h]
  \centering
  \caption{Performance comparison (I–III): Petster-hamster, Tv-show, Politician; Advogato, Public, Epinions; Twitter and Weibo. Best is highlighted in bold.}
  \fontsize{7}{8}\selectfont
  \label{tab:performance-framework-all3}
  \renewcommand{\arraystretch}{0.9}

  % ===== Block I =====
  {\normalsize
  \resizebox{\textwidth}{!}{%
    \begin{tabular}{l*{15}{c}}
      \toprule
        & \multicolumn{5}{c}{Petster-hamster}
        & \multicolumn{5}{c}{Tv-show}
        & \multicolumn{5}{c}{Politician}\\
      \cmidrule(lr){2-6}\cmidrule(lr){7-11}\cmidrule(lr){12-16}
      Methods (budget)
        & 10 & 20 & 30 & 40 & 50
        & 10 & 20 & 30 & 40 & 50
        & 10 & 20 & 30 & 40 & 50\\
      \midrule
      Random        &  35.9 &  38.1 &  92.2 & 130.8 &  94.2 &  23.8 &  64.5 &  91.5 &  79.5 & 172.2 &  52.6 &  41.2 &  74.7 & 175.2 & 124.4\\
      PageRank      &  12.9 &  45.2 &  64.7 &  75.6 &  94.2 &  17.4 &  27.4 &  43.4 &  53.8 &  63.7 &  17.7 &  28.4 &  90.6 & 105.4 & 120.3\\
      \midrule
      gIM           & 151.6 & 220.4 & 249.8 & 261.6 & 274.3 &  82.3 & 165.9 & 228.1 & 264.8 & 361.0 & 508.5 & 649.8 & 923.5 &1128.6 &1204.9\\
      S2V-DQN       & 168.0 & 177.8 & 231.5 & 232.9 & 274.2 & 164.3 & 191.0 & 234.0 & 274.4 & 300.8 & 160.2 & 756.3 &1,044.6 &1,082.5 &1,303.5\\
      ToupleGDD    & 417.7 & 442.8 & 423.4 & 483.0 & 495.1 & 533.7 & 811.3 & 988.4 &1,081.8 &1,140.4 &2,581.2 &2,762.7 &2,975.5 &3,010.1 &3,032.3\\
      DeepIM        & 360.4 & 422.2 & 463.5 & 483.7 & 503.5 & 630.8 & 749.2 & 780.5 & 1,030.5 & 1,102.3 & 2,622.0 &2,805.5 &3,028.7 &3189.2 &3,312.1\\
      BIGDN         & 381.4 & 416.7 & 450.7 & 450.2 & 468.8 & 751.6 &1,114.6 &1,208.6 &1,362.4 &1,424.3 &2,627.1 &2,870.5 &3,063.7 &3,176.2 &3,251.3\\
      \midrule
      SP-GCRL        & \textbf{429.0} & \textbf{467.6} & \textbf{485.2} & \textbf{501.4} & \textbf{542.6} & \textbf{831.0} & \textbf{1,186.4} & \textbf{1,267.4} & \textbf{1,452.2} & \textbf{1,511.3} &\textbf{2,987.6} &\textbf{3,102.8} &\textbf{3,178.6} &\textbf{3,245.7} &\textbf{3,329.0}\\
      \bottomrule
    \end{tabular}}}

  \vspace{2pt}

  % ===== Block II =====
  {\normalsize
  \resizebox{\textwidth}{!}{%
    \begin{tabular}{l*{15}{c}}
      \toprule
        & \multicolumn{5}{c}{Advogato}
        & \multicolumn{5}{c}{Public}
        & \multicolumn{5}{c}{Epinions}\\
      \cmidrule(lr){2-6}\cmidrule(lr){7-11}\cmidrule(lr){12-16}
      Methods (budget)
        & 10 & 20 & 30 & 40 & 50
        & 10 & 20 & 30 & 40 & 50
        & 10 & 20 & 30 & 40 & 50\\
      \midrule
      Random        &  28.8 & 116.0 & 216.0 & 102.6 & 181.9
                    &  17.3 &  48.7 & 111.9 & 118.1 & 251.6
                    &  20.8 &  58.4 & 134.3 & 141.7 & 301.9 \\[2pt]
      PageRank      &  95.7 & 166.4 & 286.8 & 372.7 & 434.4
                    &  26.0 &  58.5 &  68.8 &  93.5 & 106.7
                    &  31.2 &  70.2 &  82.6 & 112.2 & 228.0 \\[2pt]
      \midrule
      gIM           &2,845.7 &2,855.7 &2,865.6 &2,870.2 &2,874.3
                    & 904.6 &1,069.7 &2,039.1 &2,245.1 &2,291.4
                    & 597.4 & 701.3 &1,350.9 &1,493.4 &1,510.7 \\[2pt]
      S2V-DQN       &3,669.4 &3,669.0 &3,668.5 &36,69.0 &3,666.4
                    & 785.1 & 977.8 &1,535.4 &1,530.3 &1,578.2
                    &1,179.8 &1,179.7 &1,201.9 &1,605.0 &2,157.1 \\
      ToupleGDD    &3,670.1 &3,672.6 &3,673.3 &3,676.4 &3,681.0
                    &5,397.8 &5,440.4 &5,502.9 &5,583.6 &5,705.1
                    &3,526.1 &3,585.1 &3,658.8 &3,659.8 &3,696.7 \\
      DeepIM        &3,674.1 &3,679.4 &3,681.1 &3,686.4 &3,689.6
                    &1,600.0 &2,272.5 &3,779.3 &3,854.9 &3,879.7
                    &3,570.3 &3,607.1 &3,752.3 &3,817.0 &3,899.2 \\
      BIGDN         &3,669.4 &3,669.1 &3,669.4 &3,668.9 &3,668.1
                    &5,348.9 &5,550.7 &5,670.6 &5,726.7 &5,940.7
                    &3,506.3 &3,612.4 &3,732.5 &3,876.6 &3,960.8 \\
      \midrule
      SP-GCRL        &\textbf{3,702.1} &\textbf{3,698.0} &\textbf{3710.7} &\textbf{3,713.5} &\textbf{3,696.4}
                    &\textbf{5,514.6} &\textbf{5,766.5} &\textbf{5,820.2} &\textbf{5,932.2} &\textbf{6,035.7}
                    &\textbf{3,641.7} &\textbf{3,780.6} &\textbf{3,856.0} &\textbf{3,946.0} &\textbf{3,979.2} \\
      \bottomrule
    \end{tabular}}}

  \vspace{2pt}

  % ===== Block III =====
  {
  \resizebox{\textwidth}{!}{%
    \begin{tabular}{l*{10}{c}}
      \toprule
        & \multicolumn{5}{c}{Twitter}
        & \multicolumn{5}{c}{Weibo}\\
      \cmidrule(lr){2-6}\cmidrule(lr){7-11}
      Methods (budget)
        & 10 & 20 & 30 & 40 & 50
        & 10 & 20 & 30 & 40 & 50\\
      \midrule
      Random
        & 28,886 & 116,348 & 216,648 & 102,908 & 182,446
        &  2,699 &   7,597 &  17,456 &  18,424 &  39,250 \\[2pt]

      PageRank
        &  95,987 & 166,899 & 287,660 & 373,818 & 435,703
        &   4,056 &   9,126 &  10,733 &  14,586 &  16,645 \\[2pt]
      \midrule
      gIM
        & 2,850,943 & 2,880,517 & 2,910,268 & 2,940,733 & 2,970,184
        & 140,736  & 170,284  & 300,912  & 340,587  & 360,941 \\[2pt]

      S2V\text{-}DQN
        & 3,380,642 & 3,401,213 & 3,423,567 & 3,446,118 & 3,478,905
        & 120,438  & 150,217  & 235,906  & 242,511  & 249,873 \\

      BIGDN
        & 3,452,831 & 3,472,409 & 3,503,276 & 3,534,882 & 3,566,341
        & 824,728  & 865,903  & 896,447  & 914,205  & 936,582 \\

      ToupleGDD
        & 3,521,764 & 3,542,395 & 3,563,981 & 3,594,257 & 3,621,803
        & 782,315  & 813,604  & 842,997  & 861,732  & 883,419 \\

      DeepIM
        & 3,602,417 & 3,623,058 & 3,645,221 & 3,666,774 & 3,688,932
        & 241,368  & 332,741  & 523,518  & 582,906  & 624,177 \\

      \midrule
      SP\text{-}GCRL
        & \textbf{3,683,549} & \textbf{3,705,194} & \textbf{3,728,611} & \textbf{3,744,973} & \textbf{3,755,627}
        & \textbf{864,932}  & \textbf{905,711}  & \textbf{934,268}  & \textbf{963,754}  & \textbf{989,321} \\
      \bottomrule
    \end{tabular}}}

\end{table*}
\subsection{Impact of GAT Approximation}

We evaluate the impact of the GAT approximation to compute the subgraphs underpinning SBV (Steiner-Backbone View) and CGV (Controllability-Gramian View) on eight graphs of increasing size. Bars report end-to-end speedup when replacing the original strategy with a GAT approximation (higher is better). Lines with shaded bands show the approximation error relative to the non-approximate reference under the same view and its 95\% confidence interval. 

\begin{wrapfigure}{r}{0.5\textwidth}   % r=右侧；宽度按需调
  %\vspace{-6pt}                         % 细调与上一行间距（可删）
  \centering
  \includegraphics[width=\linewidth]{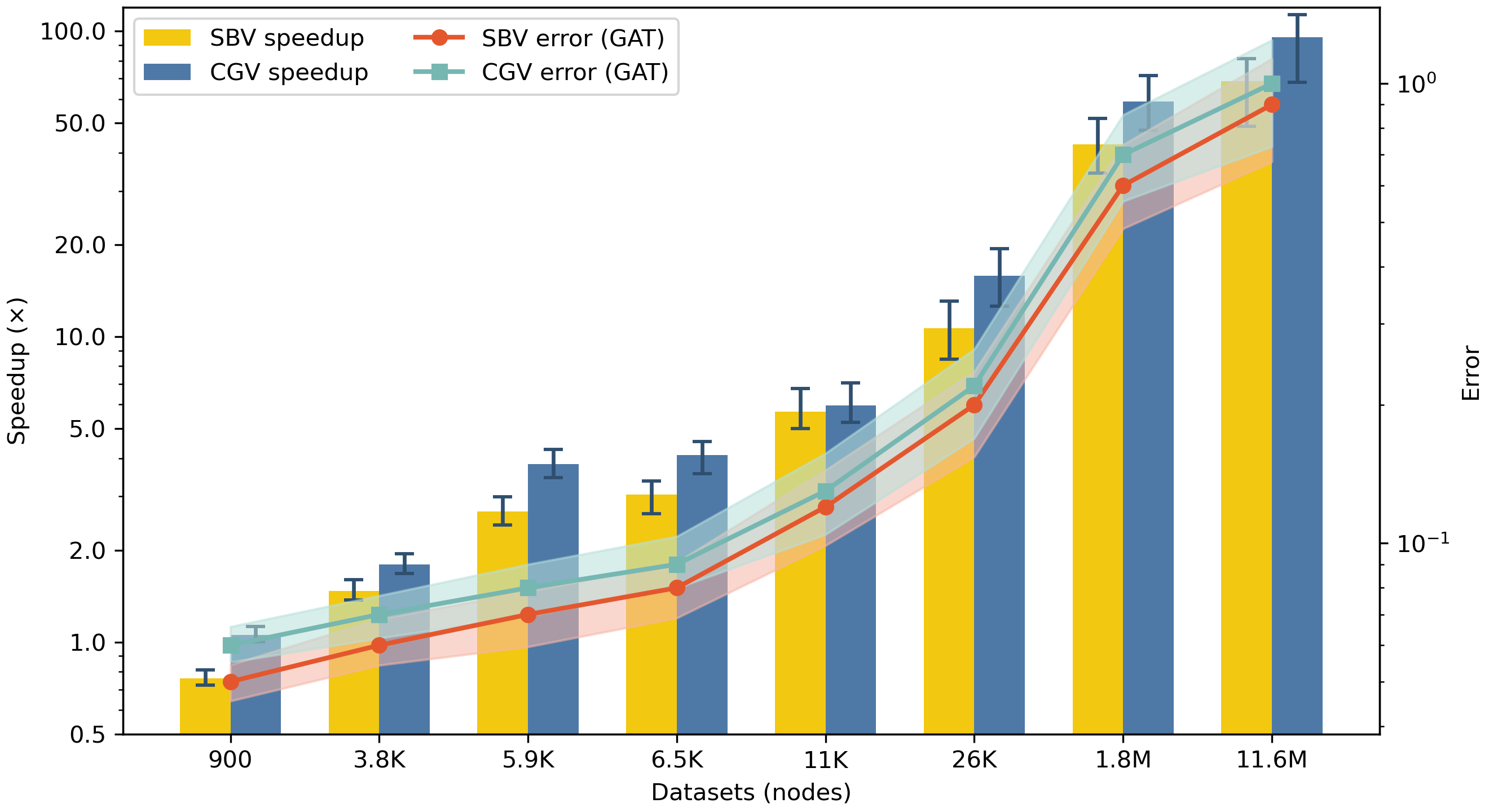}
  \caption{Speedup and error under the GAT approximation.}
  \label{fig:speedup-error}
  %\vspace{-10pt}                        % 细调与下文间距（可删）
\end{wrapfigure}

The speedup increases monotonically with graph size—from small on medium graphs (900–26K nodes) to substantial gains on large graphs (1.8M and 11.6M). Note also that the speedup obtained by the CGV approximation consistently exceeds the one obtained on SBV, with a widening gap at larger scales, suggesting more effective computational reuse and parallelism.
Errors grow mildly with size but remain acceptable; confidence bands are narrow on small graphs and broaden slightly at the million-node scale, with substantial overlap between views and no systematic bias. All results are averaged over multiple independent runs with fixed Monte-Carlo settings and identical random seeds; confidence intervals are obtained via bootstrap over run-level errors. Overall, the GAT approximation delivers scale-amplified runtime benefits for both views without noticeable accuracy loss, with CGV exhibiting the more stable efficiency advantage.
\subsection{Ablation Study}

To assess the roles of GCL and view augmentation, we construct an ablation model SP-GNN that is identical to SP-GCRL in diffusion module, reward definition, DDQN decision head, and training configuration, but replaces the encoder with an $L$-layer GNN and removes both structural view augmentations and the local–global contrastive loss. As shown in \Cref{fig:ablation_study}, SP-GCRL achieves larger diffusion across all budgets and the advantage widens with increasing budget: on Tv-show and Petster-hamster the gain is about 3–5\%; on Politician it reaches roughly twenty percent at budget 50; and on Advogato both methods fluctuate around $3.68$–$3.72\times 10^{3}$ yet SP-GCRL maintains a 1\% edge at most points with a notable jump at budget 30. The shaded bands indicate that these improvements are stable under randomness.
\begin{figure}[h]
  \centering
  % ---- 关键：把两块放进同一行，不给它换行的机会 ----
  \makebox[\linewidth][c]{%
    \begin{minipage}[t]{0.485\linewidth}
      \centering
      \subcaptionbox{\scriptsize Tv-show}[.48\linewidth]{\includegraphics[width=\linewidth]{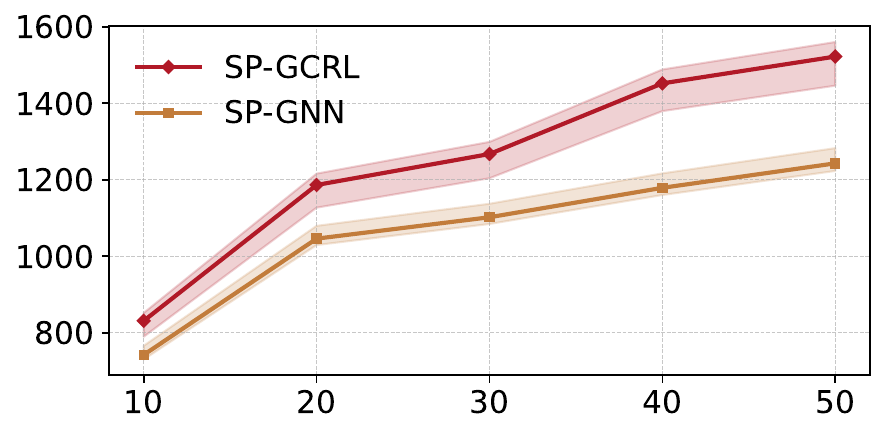}}\hfill
      \subcaptionbox{\scriptsize Politician}[.48\linewidth]{\includegraphics[width=\linewidth]{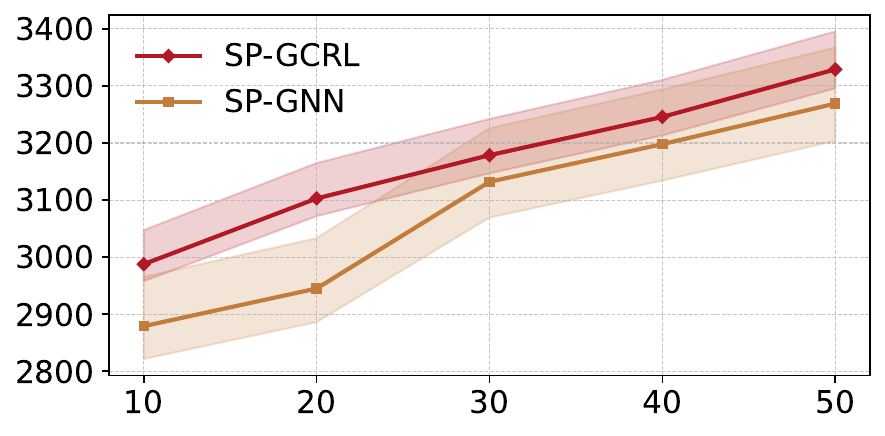}}\\[4pt]
      \subcaptionbox{\scriptsize Petster-hamster}[.48\linewidth]{\includegraphics[width=\linewidth]{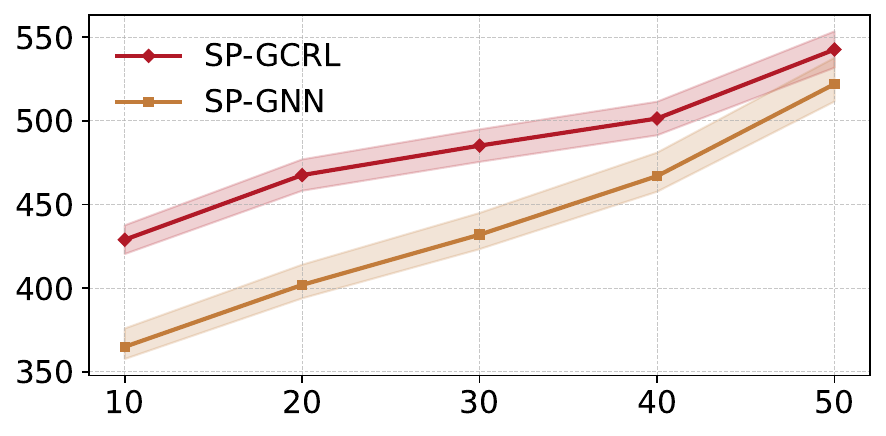}}\hfill
      \subcaptionbox{\scriptsize Advogato}[.48\linewidth]{\includegraphics[width=\linewidth]{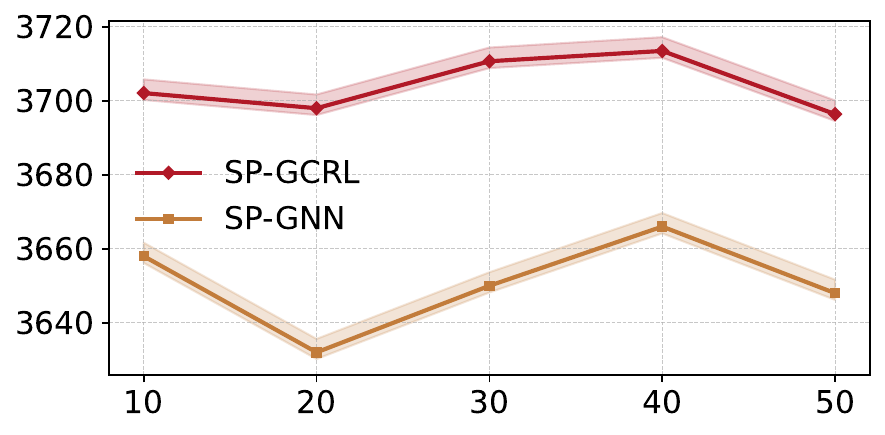}}
      \captionof{figure}{Comparison of SP-GCRL and SP-GNN on four networks.}
      \label{fig:ablation_study}
    \end{minipage}%
    \hspace{0.02\linewidth}% 中间留一点空
    \begin{minipage}[t]{0.485\linewidth}
      \centering
      \subcaptionbox{\scriptsize SP-GCRL}[.31\linewidth]{\includegraphics[width=\linewidth]{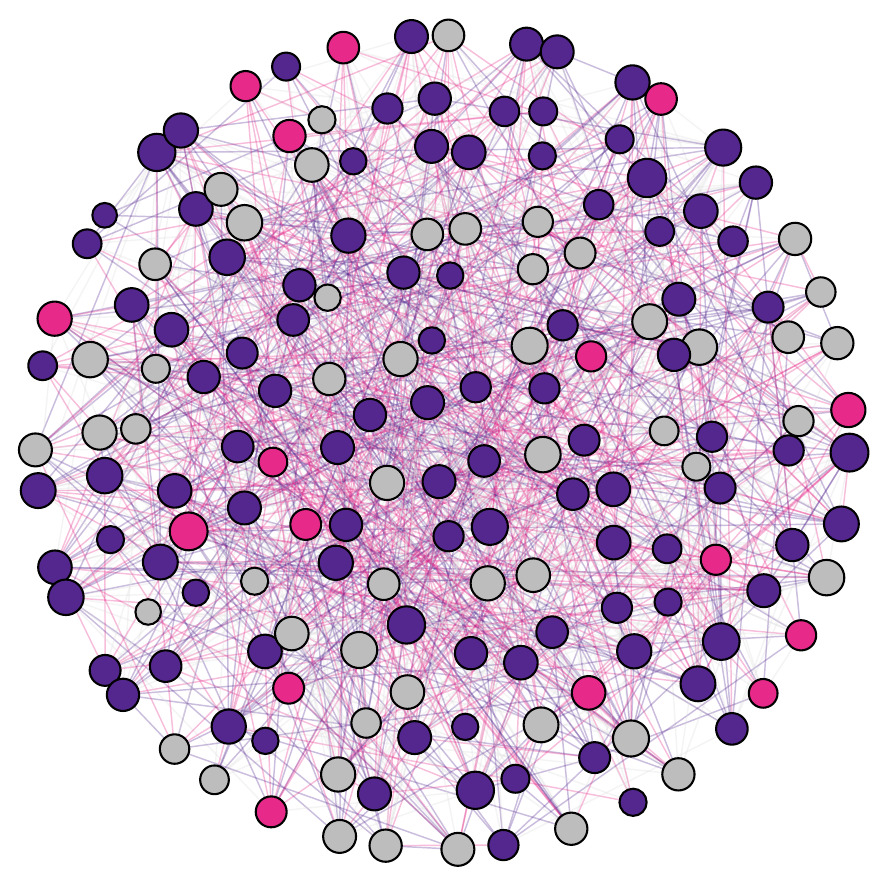}}\hfill
      \subcaptionbox{\scriptsize BIDGN}[.31\linewidth]{\includegraphics[width=\linewidth]{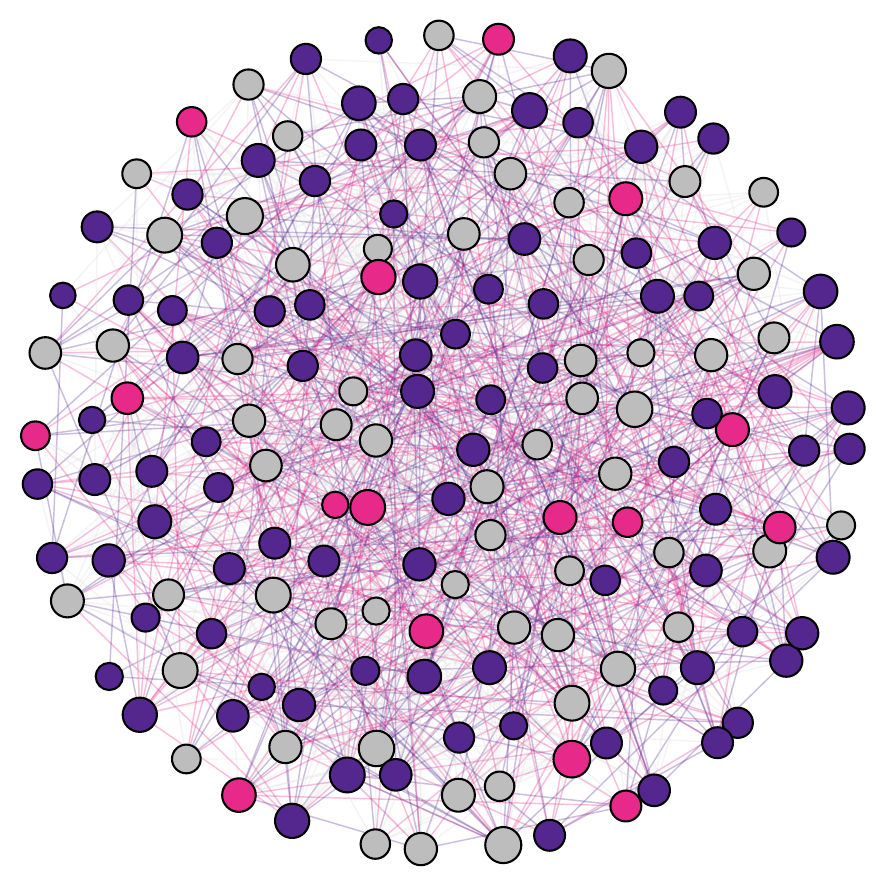}}\hfill
      \subcaptionbox{\scriptsize DeepIM}[.31\linewidth]{\includegraphics[width=\linewidth]{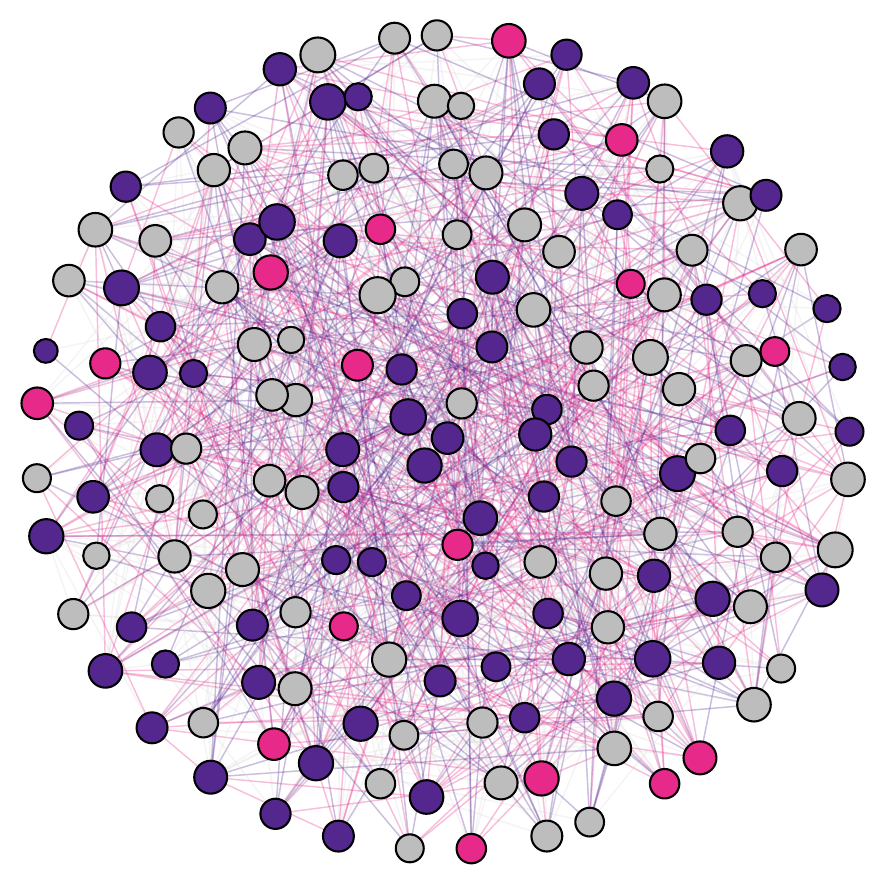}}\\[4pt]
      \subcaptionbox{\scriptsize Tou-GDD}[.31\linewidth]{\includegraphics[width=\linewidth]{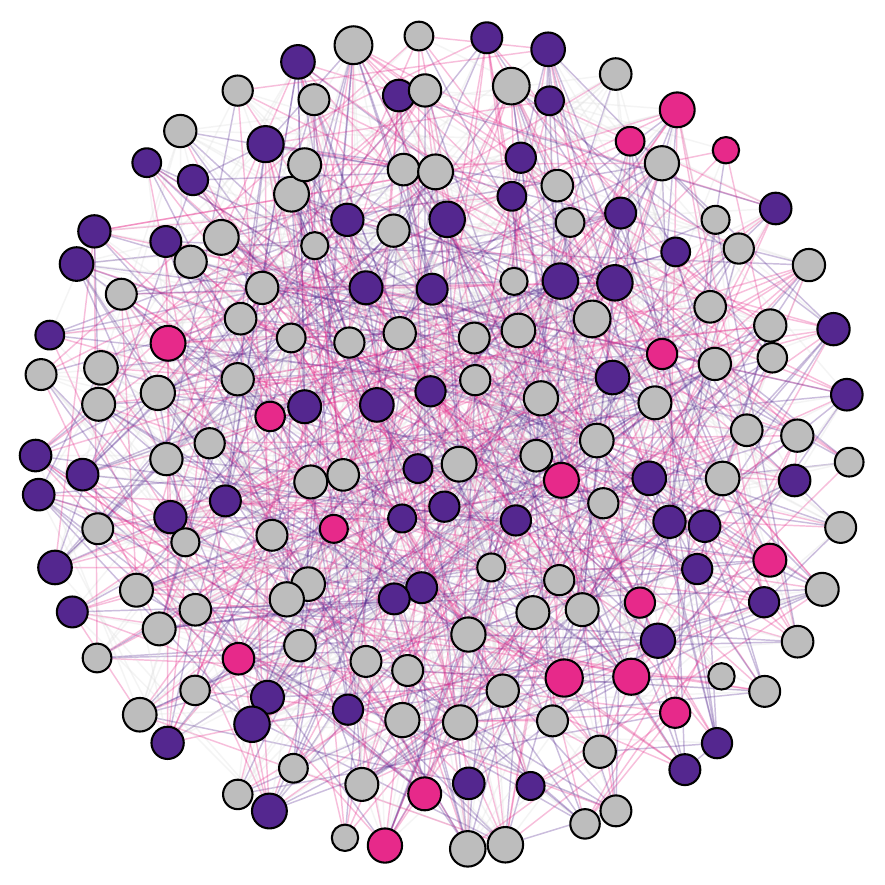}}\hfill
      \subcaptionbox{\scriptsize S2v-DQN}[.31\linewidth]{\includegraphics[width=\linewidth]{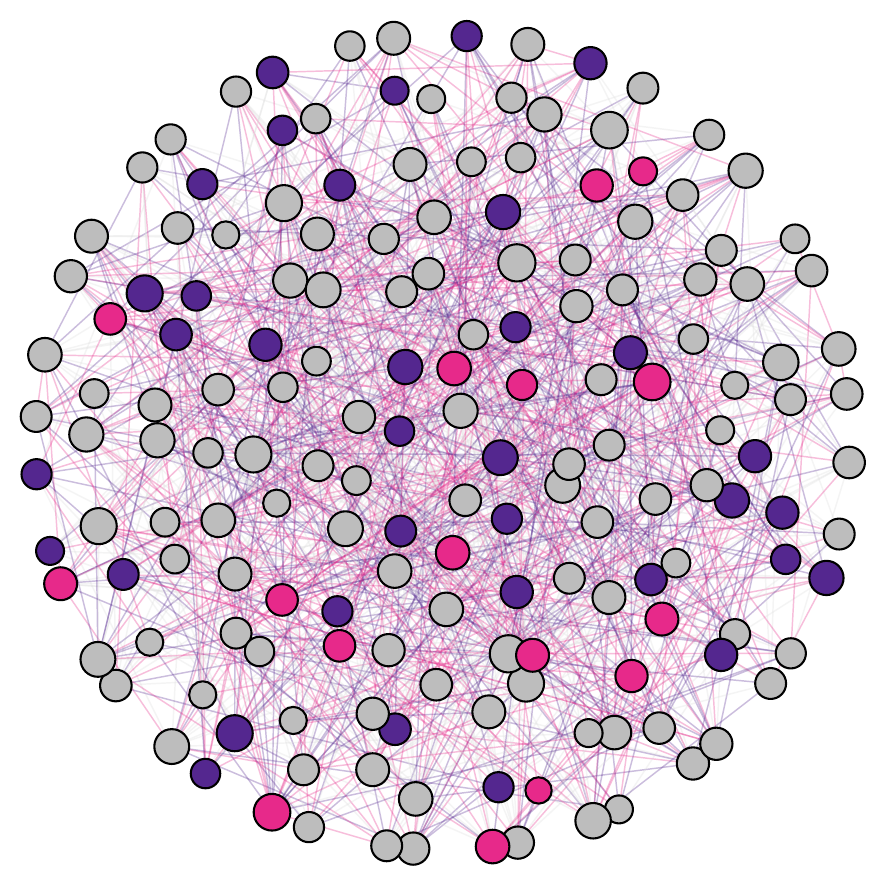}}\hfill
      \subcaptionbox{\scriptsize gIM}[.31\linewidth]{\includegraphics[width=\linewidth]{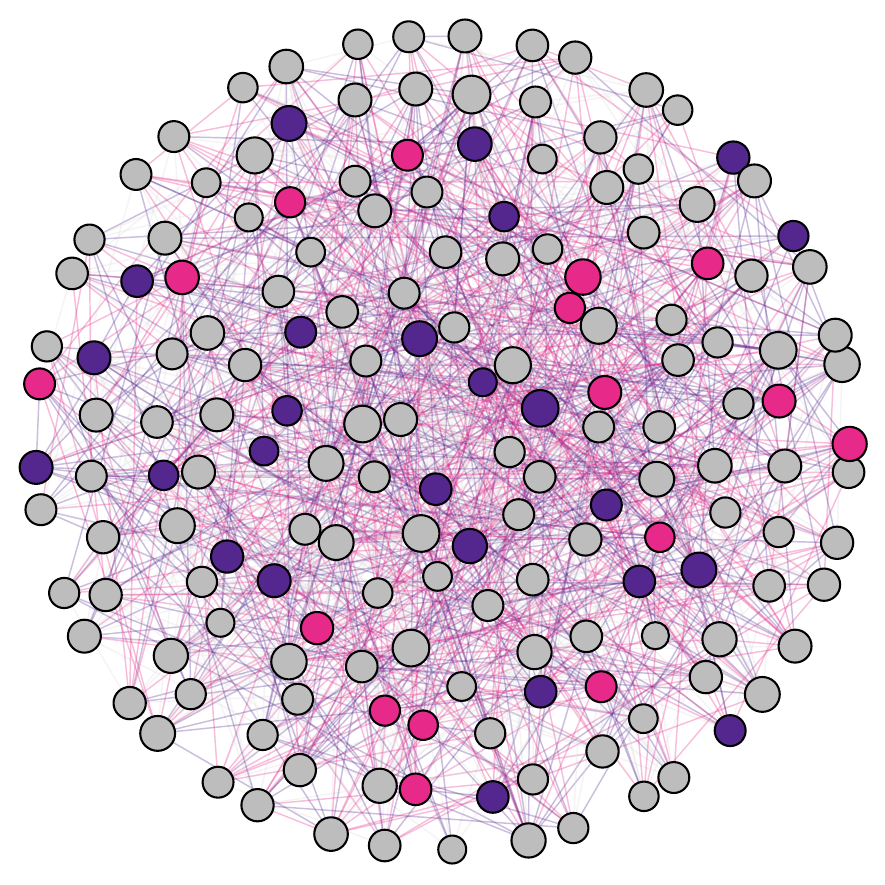}}
      \captionof{figure}{Visualization on radoslaw-email (red: seeds; blue: influenced; grey: unaffected).}
      \label{fig:case_radoslaw}
    \end{minipage}%
  }% makebox 结束
\end{figure}

\iffalse
\begin{figure}[t!]
	\centering
	\subfloat[Tv-show]{\includegraphics[width=.25\columnwidth]{Generalization_tv.pdf}}
	\subfloat[Politician]{\includegraphics[width=.25\columnwidth]{Generalization_pol.pdf}}
	\subfloat[Petster-hamster]{\includegraphics[width=.25\columnwidth]{Generalization_ph.pdf}}
    \subfloat[Advogato]{\includegraphics[width=.25\columnwidth]{Generalization_adv.pdf}}
        
	\caption{Comparison of SP-GCRL and SP-GNN on four networks. }
    \label{fig:ablation_study}
\end{figure}
\fi
\subsection{Graph Diffusion Visualization}
Finally, we conduct a case study to demonstrate the distribution of a selected set of seed nodes as well as the final propagation status of all nodes in \Cref{fig:case_radoslaw}, where red nodes indicate the initial seed node, blue nodes indicate the infected node during the influence spread, and grey nodes represent uninfected nodes. Visually, SP-GCRL disperses seeds more evenly and achieves the densest, most homogeneous cascade, whereas BIDGN, DeepIM and ToupleGDD cluster seeds in local communities, S2V-DQN and gIM leave sizeable peripheral regions uninfected, resulting in sparser overall spread.

\iffalse
\begin{figure}[htbp]
    \centering
    \subfloat[SP-GCRL]{\includegraphics[width=0.15\columnwidth]{graph_notoverlap_sized_bigdn.pdf}}\hspace{2pt}
    \subfloat[BIDGN]{\includegraphics[width=0.15\columnwidth]{graph_notoverlap_sized_touplegdd.pdf}}\hspace{2pt}
    \subfloat[DeepIM]{\includegraphics[width=0.15\columnwidth]{graph_notoverlap_sized_deepim.pdf}}\\
    \subfloat[ToupleGDD]{\includegraphics[width=0.15\columnwidth]{graph_notoverlap_sized_s2v.pdf}}\hspace{2pt}
    \subfloat[S2v-DQN]{\includegraphics[width=0.15\columnwidth]{graph_notoverlap_sized_hpc.pdf}}\hspace{2pt}
    \subfloat[gIM]{\includegraphics[width=0.15\columnwidth]{graph_notoverlap_sized_pr.pdf}}
    \caption{Visualization of influence spread on the radoslaw-email dataset. Red: seed nodes; blue: influenced nodes; grey: unaffected nodes.}
    \label{fig:case_study}
\end{figure}
\fi

\section{Conclusion}
In this work, we present SP-GCRL for influence maximization on social graphs with missing edges and noisy features. The framework combines a data-fitting propagation function, two structure-view augmentations for contrastive representation learning, a lightweight GAT surrogate that replaces costly subgraph construction, and a Double-DQN policy for seed selection. Evaluations on real networks from thousands to millions of nodes show that SP-GCRL consistently improves expected spread while reducing computation and memory. These results indicate that SP-GCRL is a practical and scalable solution for influence maximization on large, imperfect graphs.

\subsubsection{\ackname} This research is supported by the National Laboratory of Space Intelligent Control (No. HTKJ2023KL502003) and the Key Laboratory of Intelligent Space TTC\&O (Space Engineering University), Ministry of Education (No. CYK2025-01-07).

\subsubsection{\discintname}
The authors have no competing interests to declare that are
relevant to the content of this article.

\bibliographystyle{splncs04}
\bibliography{references}

\end{document}